\documentclass[11pt,a4paper]{quantumarticle}
\pdfoutput=1

\usepackage{amsmath}
\usepackage{amssymb}
\usepackage{booktabs}
\usepackage{graphicx}
\usepackage[english]{babel}
\usepackage{siunitx}
\usepackage{xcolor}
\usepackage{hyperref}

\newcommand\nn{Y}

\newcommand\layer{L}
\newcommand\half{\mbox{\small $\frac{1}{2}$}}
\newcommand\quarter{\mbox{\small $\frac{1}{4}$}}

\definecolor{comment}{gray}{0.4}

\begin{document}

\title{Completely Quantum Neural Networks}
%\preprint{}

\author{Steve Abel}
\email{steve.abel@durham.ac.uk}
\affiliation{Institute for Particle Physics Phenomenology, Durham University, Durham DH1 3LE, UK}
\affiliation{Department of Mathematical Sciences, Durham University, Durham DH1 3LE, UK}

\author{Juan C. Criado}
\email{juan.c.criado@durham.ac.uk}
\affiliation{Institute for Particle Physics Phenomenology, Durham University, Durham DH1 3LE, UK}
\affiliation{Department of Physics,  Durham University, Durham DH1 3LE, UK}

\author{Michael Spannowsky}
\affiliation{Institute for Particle Physics Phenomenology, Durham University, Durham DH1 3LE, UK}
\affiliation{Department of Physics, Durham University, Durham DH1 3LE, UK}
\email{michael.spannowsky@durham.ac.uk}

\maketitle

\begin{abstract}

Artificial neural networks are at the heart of modern deep learning algorithms. We describe how to embed and train a general neural network in a quantum annealer without introducing any classical element in training. To implement the network on a state-of-the-art quantum annealer, we develop three crucial ingredients:  binary encoding the free parameters of the network, polynomial approximation of the activation function, and reduction of binary higher-order polynomials into quadratic ones. Together, these ideas allow encoding the loss function as an Ising model Hamiltonian. The quantum annealer then trains the network by finding the ground state. We implement this for an elementary network and illustrate the advantages of quantum training: its consistency in finding the global minimum of the loss function and the fact that the network training converges in a single annealing step, which leads to short training times while maintaining a high classification performance. Our approach opens a novel avenue for the quantum training of general machine learning models.

\end{abstract}

\flushbottom

%%%%%%%%%%%%%%%%%%%%%%%%%%%%%%%%%%%%%%%%%%%%%%%

\section{\label{Sec:Intro}Introduction}

Neural networks (NN) have become an important  machine learning tool, in particular for classification tasks, and there is great interest in improving their performance using quantum computing techniques.
Indeed a neural network contains three essential features that one might seek to enhance this way, namely
\begin{enumerate}
\item an adaptable system that approximately encodes a complicated function,
\item a loss function in the output layer whose minimisation defines the task the NN algorithm should perform,
\item a training algorithm that minimizes the loss function.
\end{enumerate}
To date, it has been possible to enhance one or more of these three aspects using quantum computing.
For example the encoding in (1) can be implemented on quantum gate and continuous variable quantum devices~\cite{McClean_2016,farhi2018classification,Killoran_2019_NN,PhysRevA.101.032308,Blance:2020ktp,mari2020transfer,Ngairangbam:2021yma,koji2021,eisert2021,Araz:2022haf}.
The loss function can also be implemented in an entirely quantum fashion~\cite{Peruzzo14,farhi2014quantum,Blance:2020nhl} (as a function of the network outputs), and the minimisation of the loss function can be implemented on quantum devices in several ways, in particular using quantum annealers~\cite{farhi2000quantum,nevenNIPS2009,gorman2015,Mott:2017xdb,Zlokapa:2019lvv,sasdelli2021quantum}.
The result has been hybrid quantum/classical implementations that have nevertheless often demonstrated some improvement.
However, it has to date not been possible to encode all three of these aspects, in other words, to implement an entire neural network algorithm onto a single quantum device with no classical elements at all. 

The purpose of this work is to implement such a completely quantum neural network to solve the task of binary classification of certain well known (and not so well known) data sets and investigate how it compares with classical devices. 

Our purpose in this implementation is somewhat different from previous studies. We envisage the neural network conventionally as simply a gigantic function that we wish to optimise during a training phase by adjusting the weights and biases in the network. Our objective is to achieve this training in a quantum way in one step by encoding the network in its entirety on a quantum device. Thus our implementation must incorporate several aspects that have not to date been combined. Of primary importance are the following. First, the network must include non-trivial activation functions for the weights and biases. These activation functions, along with the weights and biases, must somehow be incorporated into the network by encoding onto the quantum device.
Conversely, however, to be effective and competitive, the training stage must be able to utilise virtually unlimited data (or at least a data set that can easily be larger than a number of qubits on any device currently in existence). This means that one should avoid trying to encode the data itself onto the quantum device (otherwise, a one-step training would not be possible) but should include all the data directly in the loss function. We will produce practically applicable neural networks by paying attention to these two essential requirements.

The device we utilize for this task is a quantum annealer~\cite{finilla94a,kadowaki98a,brooke99a,dickson13a,lanting14a,albash15a,albash16a,boixo16a,chancellor16b,Benedetti16a,Muthukrishnan2016,cervera,LantingAQC2017}, that choice being dictated solely by the large numbers of qubits required to encode the neural network. Indeed as we shall see, the main limitations of the method are the number of qubits required to encode the network itself, which restricts the number of features, and the number and size of hidden layers. However, even working within these restrictions we will be able to show that the method works effectively and to illustrate the two main advantages of the quantum training: that it can consistently find the global minimum of the loss function and that this can be done in a single training step, as opposed to the iterative procedures commonly used classically. The general method we describe here will be applicable more generally as the technology develops.

\section{\label{Sec:QA}Quantum Annealing and the Ising Model Encoding}

Let us begin with a brief description of the device, and the central features for this study. Generally a quantum annealer performs a restricted set of operations on a quantum system described by: a Hilbert space that is the tensor product of several 2-dimensional Hilbert spaces (i.e the qubits) and a Hamiltonian of the form
\begin{align}
  \mathcal{H}(s) &~=~ A(s) \sum_\ell \sigma_{\ell,x}\\
    &\qquad+ \, B(s) \left(\sum_\ell h_\ell \sigma_{\ell,z} + \sum_{\ell m} J_{\ell m} \sigma_{\ell,z} \sigma_{m,z}\right) \, , \nonumber 
\end{align}
where $\sigma_{\ell,x}$ and $\sigma_{\ell,z}$ are the corresponding Pauli matrices acting on the $\ell^{\rm th}$ qubit, and $A(s)$, $B(s)$ are smooth functions such that $A(1) = B(0) = 0$ and $A(0) = B(1) = 1$, which are used to change the Hamiltonian during the anneal.
The annealer can perform the following operations:
\begin{itemize}
  \item Set an initial state that is either the ground state of $\mathcal{H}(0)$ (known as \emph{forward annealing}) or any eigenstate of $\bigotimes_\ell \sigma_{\ell, z}$ (\emph{backward annealing}).
  \item Fix the internal parameters $h_\ell$ and $J_{\ell m}$ of the Hamiltonian $\mathcal{H}(s)$.
  \item Allow the system to evolve quantum-mechanically while controlling $s$ as a piecewise-linear function $s(t)$ of time $t$, with $s(t_{\rm final}) = 1$, and $s(t_{\rm init}) = 0$ for forward annealing, or $s(t_{\rm init}) = 1$ for backward annealing, where $t_{\rm init}$ and $t_{\rm final}$ are the initial and final times. The function $s(t)$ is called the \emph{annealing schedule}.
  \item Measure the observable $\bigotimes_\ell \sigma_{\ell,z}$ at $t = t_{\rm final}$.
\end{itemize}

Typically one chooses an anneal schedule such that the machine returns the ground state of the Ising-model Hamiltonian $\mathcal{H}(1)$.
This allows one to use it to solve optimisation problems that can be formulated as minimisation of a quadratic function $H$ of spin variables $\sigma_\ell = \pm 1$:
\begin{equation}
\label{eq:isingH}
 H(\sigma_\ell) ~=~ \sum_\ell h_\ell \sigma_\ell + \sum_{\ell m} J_{\ell m} \sigma_\ell \sigma_m~.
\end{equation}
To differentiate between the physical system and the embedded abstract problem, we will refer to the elements of the physical system ${\cal H}(\sigma_{\ell,z})$ as \emph{qubits}, and to the ones of the abstract system $H(\sigma_\ell)$ as \emph{spins}. Note that the classical spin values, $\sigma_\ell$, do not carry a $z$ index. The objective is usually that the minimisation of the physical Hamiltonian ${\cal H}(\sigma_{\ell,z})$ should yield a solution to the problem encoded by the minimisation of the problem Hamiltonian $H(\sigma_\ell)$.

Two crucial practical elements will need to be tackled to proceed with the encoding of a neural network. The first is that Eq.~\eqref{eq:isingH} describes a 
generalised Ising model, but in practice, the quantum-annealing device only allows the setting of a limited number of non-vanishing couplings $J_{\ell m}$ between qubits.
 Let us be specific to the architecture we will be using in this work, namely D-Wave's~\cite{LantingAQC2017} \texttt{Advantage\_system4.1}: this annealer contains 5627 qubits, connected in a \emph{Pegasus} structure, but only has a total of 40279 couplings between them.
Ising models with a higher degree of connectivity (more couplings) must be {\it embedded} into the physical system by chaining several qubits together with large couplings between them and treating the chain as if it were a single qubit. This step is carried out by an embedding algorithm.

The second aspect that we will need to address to treat neural networks is the fact that the functions we will need to optimise (i.e. our problem Hamiltonians $H(\sigma_\ell)$) are polynomial in spins, but the 
Ising model is only quadratic. Routines to reduce polynomial spin models to quadratic ones using auxiliary qubits (such as \texttt{make\_quadratic}) do exist in the \texttt{dimod} package, but our experience with these was limited. Being able to treat higher-order problems is an essential extension to quantum annealers: therefore, in Appendix~\ref{app:higher}, we prove that the issue of minimising {\it any} higher-order polynomial in binary variables can be transformed into the minimisation problem of a generalised quadratic Ising model of the form in Eq.~\eqref{eq:isingH}, and is thus in principle solvable by a quantum annealer.

\section{\label{sec:QNN}Encoding a quantum neural network}

\subsection{Neural networks and classical training}

A Neural Network (NN) is a highly-versatile machine-learning model built as a composition of functions with a vector input and output, known as layers. Each layer consists of a linear transformation followed by element-wise application of non-linear functions $g$, known as the \emph{activation function}. The $i^{\rm th}$ component of the output of a layer $\layer$ is thus given by
\begin{equation}
\label{eq:lossgen}
 \layer_i(x) ~=~  g\left(\sum_j w_{ij} x_i + b_i\right),
\end{equation}
where the $w_{ij}$ and $b_i$ are free parameters, the so-called \emph{weights} and \emph{biases}. Each function $L_i$ of vector input and scalar output is known as a \emph{unit}. A NN is then defined by a collection of layers $L^{(k)}$ through
\begin{equation}
 \nn ~=~ L^{(n)} \circ \ldots \circ L^{(0)}.
\end{equation}
The \emph{depth} of a NN is the number of layers $n$, while its \emph{width} is the number of units per layer (or in the largest layer if the layers vary in size). The various versions of the \emph{universal approximation theorem}~\cite{cybenko1989, HORNIK1991251, lu2017expressive} ensure that NNs that are either sufficiently deep or sufficiently wide can approximate any function with arbitrary precision. This makes them suitable for general regression and classification tasks, on which they have proven to be an efficient parametrisation.

\begin{figure}
 \centering
 \includegraphics[width=0.5\textwidth]{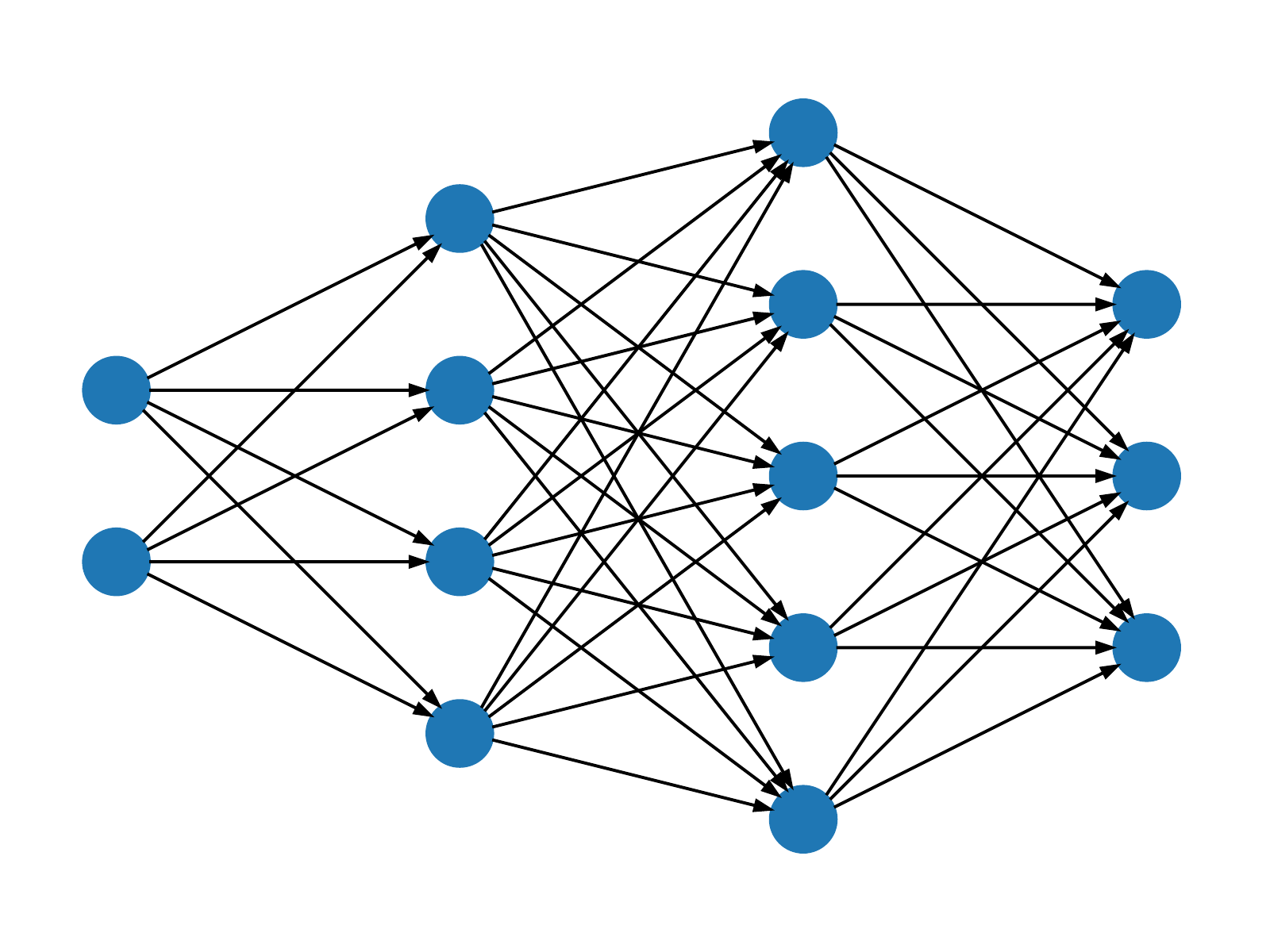}
 \caption{Schematic representation of a neural net.}
 \label{fig:net_diagram}
\end{figure}

The \emph{training} of a NN is the procedure by which its internal parameters $w^{(k)}_{ij}$ and $b^{(k)}_i$ are adjusted so that it solves the problem at hand. This is done utilising a \emph{loss function} $\mathcal{L}(Y)$, chosen such that a NN with the desired properties sits at its minimum. Typically a classical training algorithm will implement an improved version of gradient descent on $\mathcal{L}$ filled with training data to find this optimum configuration of weights and biases.

In the \emph{supervised learning} framework, the input data for the training consists of a collection of $N_d$ data points $x_a \in \mathbb{R}^{N_f}$ and a collection of the $N_d$ corresponding outputs $y_a \in \mathbb{R}^{N_o}$ to be reproduced as $y_a \simeq Y(x_a)$. The dimension $N_f$ of the input data-point space is known the \emph{number of features}. In general, when the outputs $y_a$ are arbitrary points in a vector space $\mathbb{R}^{N_o}$, the problem to be solved is a \emph{regression} one. In the particular case in which $y_a$ takes values in a discrete set, then one has a \emph{classification} problem, but the set of abstract labels can still be encoded as a set of isolated points in $\mathbb{R}^{N_o}$. For example if we seek a set of $N_o$ yes/no decisions then the outputs live in $y_a \in (\mathbb{Z}_2)^{N_o}\subset \mathbb{R}^{N_o}$. Thus both kinds of problem can be treated within the general framework described here. 

A typical loss function for supervised learning is the MSE for the outputs:
\begin{equation}
  \mathcal{L}(Y) ~=~ \frac{1}{N_d} \sum_a \left|y_a - Y(x_a)\right|^2~.
  \label{eq:MSE}
\end{equation}
This is widely used for general regression problems. It is also a viable candidate for classification, although, depending on the training method, other loss functions, such as the binary/categorical cross entropy, can be more effective.

\subsection{Training a NN in a quantum annealer}
\label{subsec:training}

Let us now consider the task at hand, namely how to encode and train such a NN on a quantum annealer. Since the purpose of an annealer is to find the minimum of a function, the Hamiltonian, we aim to write the loss function $\mathcal{L}$ as an Ising model Hamiltonian. Then, we expect the final state of the annealing process to give the optimal NN for the problem under consideration.

The loss is ultimately a function of the internal parameters of the NN, the weights $w_{ij}$ and biases $b_i$. Meanwhile the Ising model Hamiltonian $H(\sigma_\ell)$ is a function of the Ising model spins $\sigma_\ell$. Therefore, as a first step, we need a translation between the $w_{ij}$, $b_i$ parameters and the $\sigma_\ell$ spins. It is simpler for this purpose to use a QUBO encoding, related to the
spin encoding as
\begin{equation}
\tau_{\ell}~=~\frac{1}{2}(\sigma_{\ell}+1)~,
\end{equation}
where $\tau_\ell = 0,1$. Then, each of the parameters $p \sim w^{(k)}_{ij}$, $b^{(k)}_i$ is encoded in a binary fashion,
in terms of the annealer spins as
\begin{equation}
 p ~=~ -1+ \frac{1}{1 - 2^{-N_b}} \sum_{\alpha = 0}^{N_b-1} 2^{-\alpha} \tau_\alpha^p~.
 \label{eq:weight-encoding}
\end{equation}
We will use a superindex $p$ on the $\tau$ to indicate which particular block of $N_b$ qubits (labelled by $\alpha=0\ldots N_b-1$) is being used to encode that weight or bias. The above encoding yields $p \in [-1,1]$.

\begin{figure}
  \centering
  \includegraphics[width=0.45\textwidth]{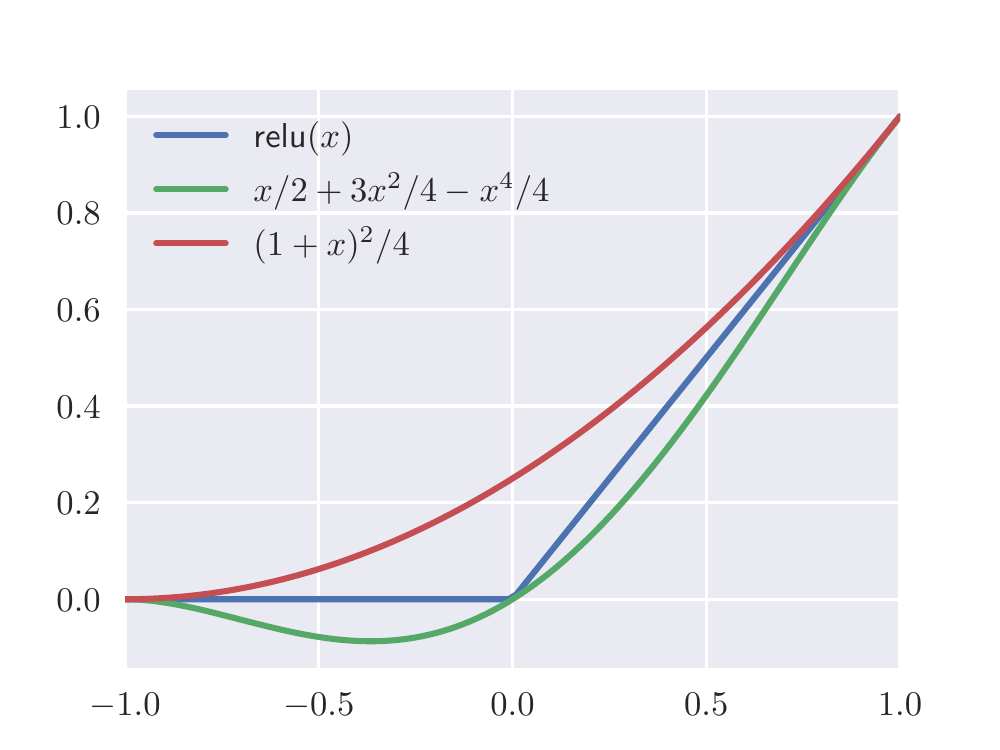}
  \includegraphics[width=0.45\textwidth]{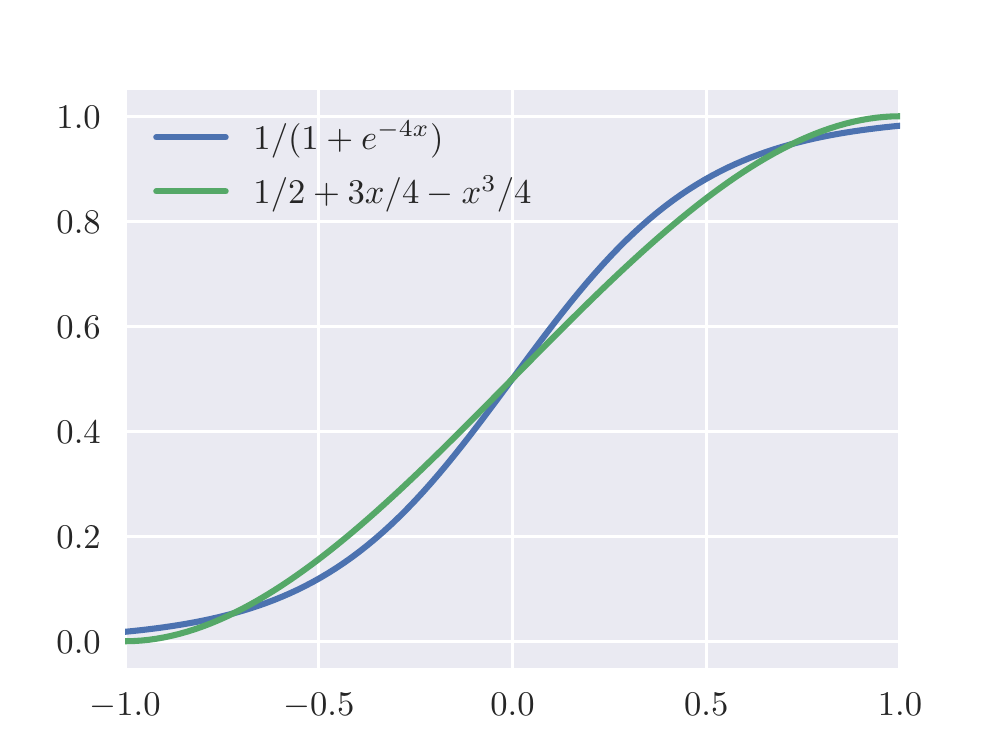}
  \caption{Polynomial approximations to the popular ReLU (top) and sigmoid (bottom) activation functions.}
  \label{fig:polynomial-activation}
\end{figure}

Using Eq.~\eqref{eq:weight-encoding}, we can write the loss as a function of the Ising model spins $\tau$. In general, this will not take the form of an Ising model Hamiltonian (defined in Eq.~\eqref{eq:isingH}), so the next step is to transform it into one. For this purpose, we first approximate the activation function by a polynomial. Since the weights and biases are bounded in this approach, the input to the activation is bounded, and therefore a polynomial can approximate it arbitrarily well in the input range. Some polynomial approximations to standard loss functions are shown in Fig~\ref{fig:polynomial-activation}.

The use of polynomials as activation functions is a delicate issue because some versions of the universal approximation theorem require the activation to be non-polynomial. In the present context, this need not concern us, however, since the boundedness of the input implies here that there is boundedness of the output, and this is enough to guarantee the universal approximation~\cite{HORNIK1991251}.

The loss function of the output value $Y(x)$ of the NN can either be a polynomial, as in Eq.~\eqref{eq:MSE}, or not. However, the above boundedness argument can admit its approximation by a polynomial even if it is not. We arrive, therefore, with the polynomial activation functions at a loss function that is a polynomial in the Ising model spins.

The final step is to then transform this polynomial into a quadratic, which would then match the form of the Ising model Hamiltonian defined in Eq.~\eqref{eq:isingH}. To do this, we may now employ the reduction method derived in Appendix~\ref{app:higher}, which can, in principle, be used to reduce a polynomial of any degree to a quadratic. 
The reduction method makes iterated use of the following quadratic polynomial in binary variables $x, y, z = 0, 1$:
\begin{equation}
  Q(z; x, y) ~=~ \Lambda (x y - 2 z (x + y) + 3 z)~.
  \label{eq:constraint-Hamiltonian}
\end{equation}
As discussed in Appendix~\ref{app:higher}, this polynomial has degenerate global minima, which are achieved for every possible value of the $(x, y)$ pair, if and only if $z = x y$, and one can check that $Q(xy;x,y)=0$ at these minima. Thus $Q$ can be used as a \emph{constraint Hamiltonian} because reaching its minimum implies that the $z = x y$ constraint is satisfied, at which point there is no nett contribution to the Hamiltonian. We can then, in principle, reduce the degree of the loss function polynomial by replacing products $xy$ of spins with auxiliary spins, $z$, and simultaneously adding $Q(z; x, y)$ to the Hamiltonian, with a sufficiently large value of $\Lambda$.  

This completes the general method for encoding the loss function into an Ising model Hamiltonian.
Before we consider specific examples let us summarise the steps we have taken in the encoding:
\begin{enumerate}
  \item Write the loss as a function of the Ising model spins by means of Eq.~\eqref{eq:weight-encoding}.
  \item Re-write it as a general polynomial in the spins by approximating the activation functions (and possibly the final operation in the loss function) by polynomials.
  \item Transform it to a quadratic polynomial by introducing auxiliary variables and adding copies of the constraint Hamiltonian defined in Eq.~\eqref{eq:constraint-Hamiltonian}.
\end{enumerate}

\subsection{An example encoding}

Let us now illustrate the procedure outlined with a concrete example. We consider a binary classification problem, with labels $y = \pm 1$. We will use a NN with a single hidden layer (two layers in total) without an activation function for the last layer to classify any input data. In detail, such a neural net produces a classification output from the inputs, which following Eq.~\eqref{eq:lossgen} is as follows:
\begin{align}
 \nn_{v,w}(x_{j}) &~=~ v_i g(w_{ij} x_{j}) + v_0~, 
\end{align}
where the $w_{ij}$ and $v_i$ are the weights for first and second layer, respectively; summation over the $i$, $j$ indices is implicit; and $i=1\ldots N_h$ labels the $N_h$ units in the hidden layer. The $x_{j}$ are assumed to contain a constant feature incorporated by adding a 0-index for biases,
so that $w_{i0}$ is the bias in the first layer, and $v_0$ is the bias in the hidden layer, while $w_{ij}$ are the weights. To reduce clutter we can 
also express the hidden layer bias $v_0$ by adding a $w_{00}$ weight as well, so that the combined system of weights and biases is encompassed by simply extending all the sums to $i=0\ldots N_h$, $i=0\ldots N_f$. With this shorthand being understood, we will therefore write 
\begin{align}
 \nn_{v,w}(x_{j}) &~=~ v_i g(w_{ij} x_{j}) ~.
\end{align}

A given state of the neural network has a certain $w$ and $v$, and the result of inputting data $x_j$ is a single output $Y_{v,w}(x_j)$ that is used to decide its classification. The prediction is $y = 1$ if $\nn_{v,w}(x) > 0$ and $y = -1$ otherwise. (Or, to put it another way, there is a final Heaviside activation function, $y=2\vartheta(Y_{v,w}(x))-1$ feeding into the binary classification $y$.)

As mentioned, the crux of the matter is now to optimise $v$ and $w$ by training the neural network on data. To perform this optimisation, we need to feed into the neural net a set of input data $x_{ai}$, where $a$ labels the data points, and to optimise the weights and biases to give the best match with the known classifications $y_a=\pm 1$ that correspond to this data. We will use Eq.~\eqref{eq:MSE} as the loss function.

For the activation function, consider the simple approximation to the ReLU function (shown in Figure~\ref{fig:polynomial-activation}) given by $g(x) = (1 + x)^2 / 4$. 
We first use Eq.~\eqref{eq:weight-encoding} to encode the weights, and then this results in a loss function that is sextic in spins. 
Since it is a total square, it is more straightforward to reduce the loss function to a quadratic by eliminating pairs of spins in
\begin{align}
Y_{v,w}(x) ~&=~ y_a - \quarter -\half v_i w_{ij} x_{aj} \nonumber \\
~&\qquad\qquad ~~ - \quarter v_i w_{ij} w_{ij'} x_{aj} x_{aj'}\,,
\label{eq:expanded-Y}
\end{align}
with the use of auxiliary variables as described in Sec.~\ref{subsec:training}, until $Y_{v, w}(x)$ becomes a linear function of the spins.

To replace the $v_i w_{ij}$ term for example, for every quadruple $\alpha i;\beta j$
we trade the pairs of binary variables that appear in the product with auxilliary variables by adding 
\begin{align}
 H_{vw }  ~&=~ \sum_{i,j
        %\stackrel{i,j}{\scriptscriptstyle \alpha+\beta\ < N_b}
        }Q( \tau^{(ij)}_{\alpha\beta}
   ;
   \tau^{\scriptscriptstyle v_i }_\alpha 
   \tau^{\scriptscriptstyle w_{ij}}_\beta ) ~.
\end{align}
This requires $$N_{vw}~=~(N_h+1) (N_f+1) N_b^2$$ auxiliary qubits, so that for example $N_h=N_f=2$ and $N_b=1$ requires only $9$ auxiliary qubits.
We may then go on to replace the terms in $v_i w_{ij}w_{ij'}$, by adding for every sextuple $\alpha i;\beta j;\gamma j'$ the constraint Hamiltonian 
 \begin{align}
  H_{vww }  ~&=~ \sum_{i,j,j'
         %\stackrel{i,j,j'}{\scriptscriptstyle \alpha+\beta+\gamma\ < N_b}
         }Q(
   \tau^{(ijj')}_{\alpha\beta\gamma}
   ;
   \tau^{(ij)}_{\alpha\beta}
   \tau^{\scriptscriptstyle w_{ij}}_\gamma ) ~,
 \end{align}
requiring a further $$N_{vww}~=~(N_h+1) (N_f+1)^2 N_b^3$$ auxiliary qubits. And so forth for higher terms in the approximated activation function if included. It should be noted that the number of qubits grows geometrically with the degree of the term being reduced.

\begin{figure*}
  \centering
  \includegraphics[width=0.49\linewidth]{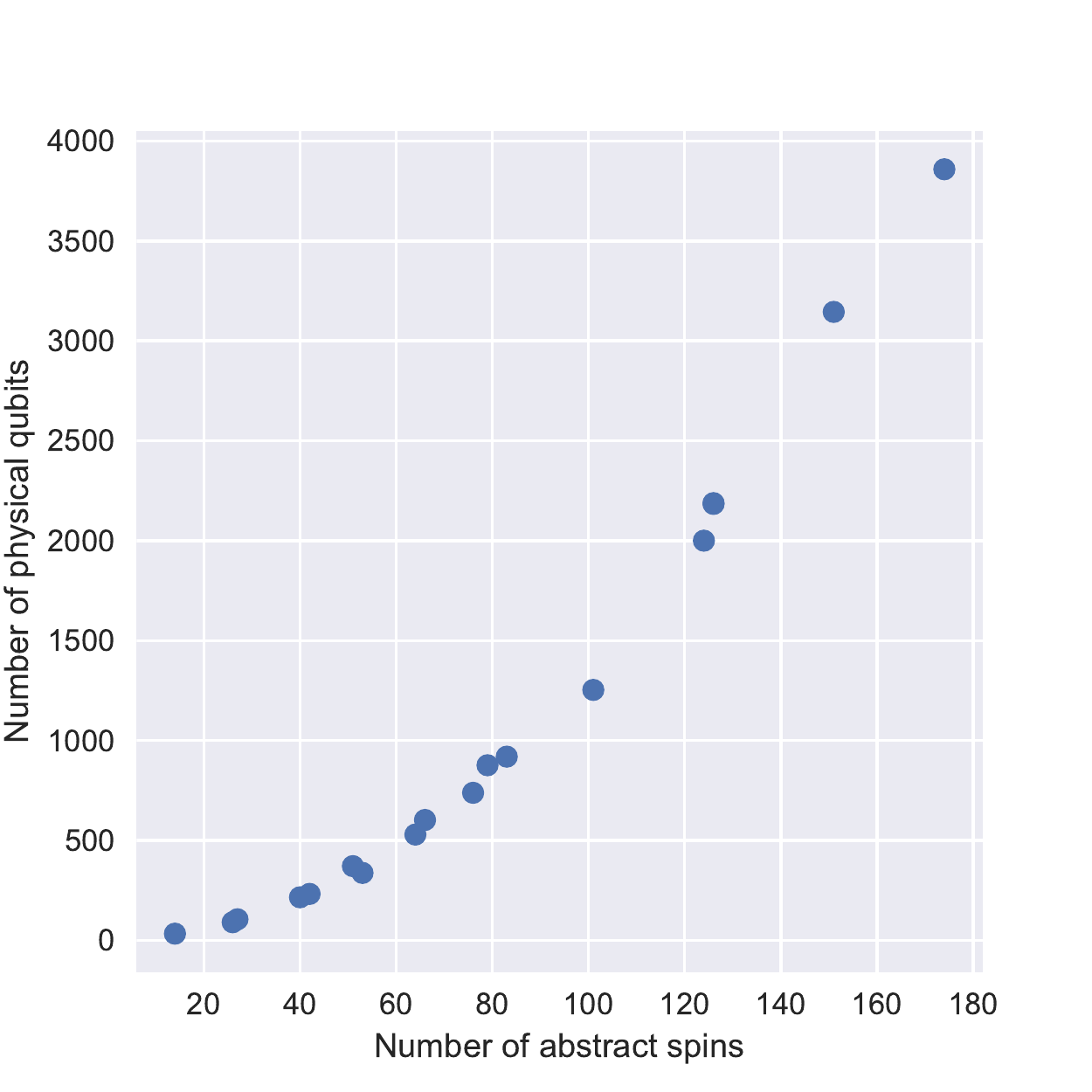}
  \includegraphics[width=0.49\linewidth]{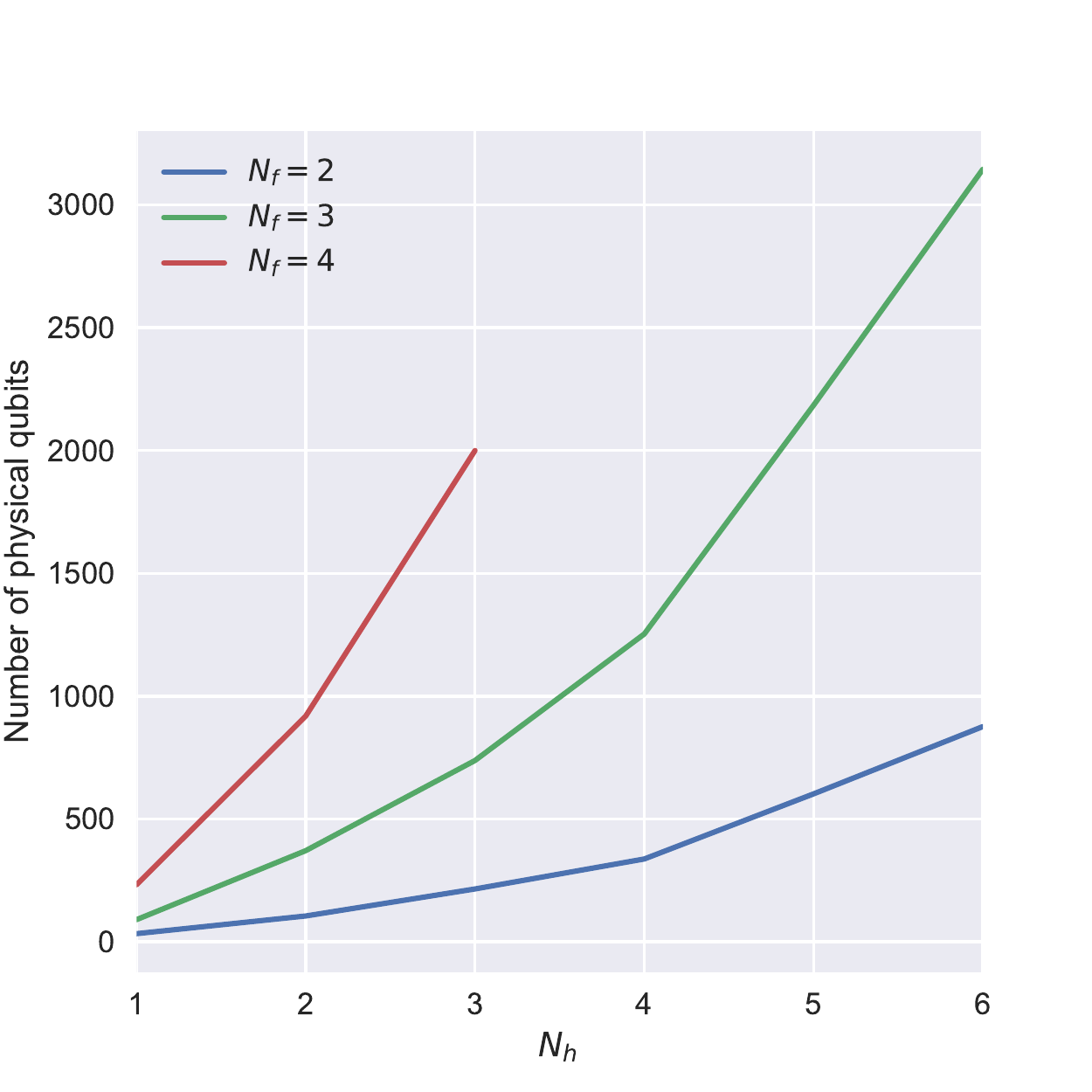}
  \caption{\label{fig:qubits}Left: number of physical qubits needed to embed the quantum NN as a function of the number of spins in its abstract encoding, for varying values of the number of features $N_f$, the number of units in the hidden layer $N_h$ and the number of spins per parameter $N_b$. Right: number of qubits as function of $N_h$, for each value of $N_f$ and fixed $N_b = 1$.}
\end{figure*}

\section{Implementation and results}

Let us now demonstrate that using the encoding discussed in Sec.~\ref{sec:QNN} one can train a neural network in a D-Wave quantum annealer. Since the encoding requires a high degree of connectivity between spins, only small networks can be currently implemented. To reduce the number of qubits needed to embed the network in the annealer, we fix the biases for the first layer to zero and set the activation function to $g(x) = x^2$. The resulting model retains all the building blocks described in Sec.~\ref{sec:QNN}: binary encoding of the network parameters, polynomial activation function, and encoding of products through a constraint Hamiltonian.

As the embedding is a function of the connectedness of the final model, and therefore rather hard to quantify, we first perform a scan over the number of features $N_f$, the number of hidden units $N_h$ and the number of spins per network parameter $N_b$, to record the number of abstract spins required by the encoding in Sec.~\ref{sec:QNN} and the corresponding number of qubits of the embedded model. The results are shown in Fig.~\ref{fig:qubits}. Networks of this kind with up to $\sim 180$ spins can thus be embedded in currently available annealers.

\begin{figure*}
 \centering
 \includegraphics[width=0.49\textwidth]{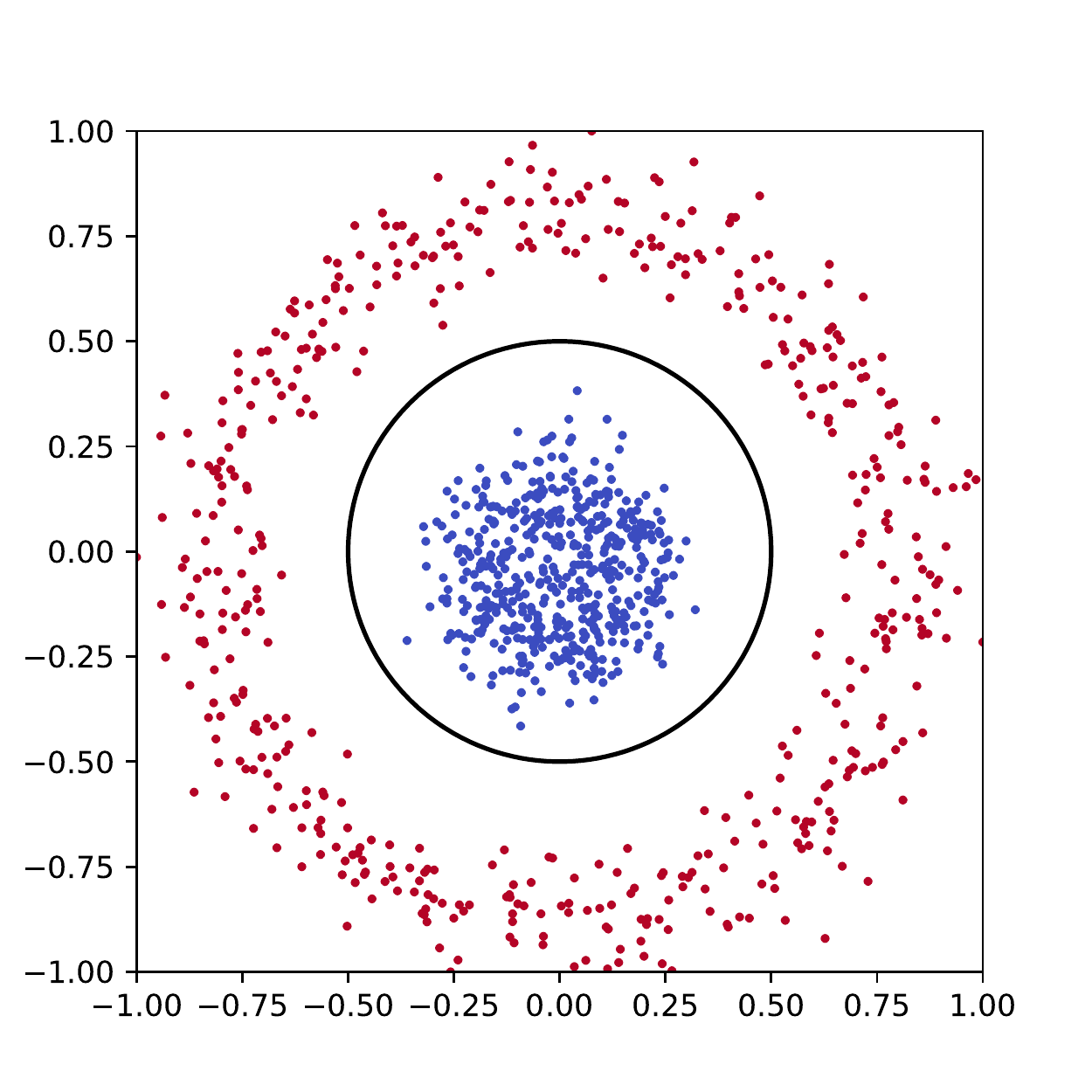}
 \includegraphics[width=0.49\textwidth]{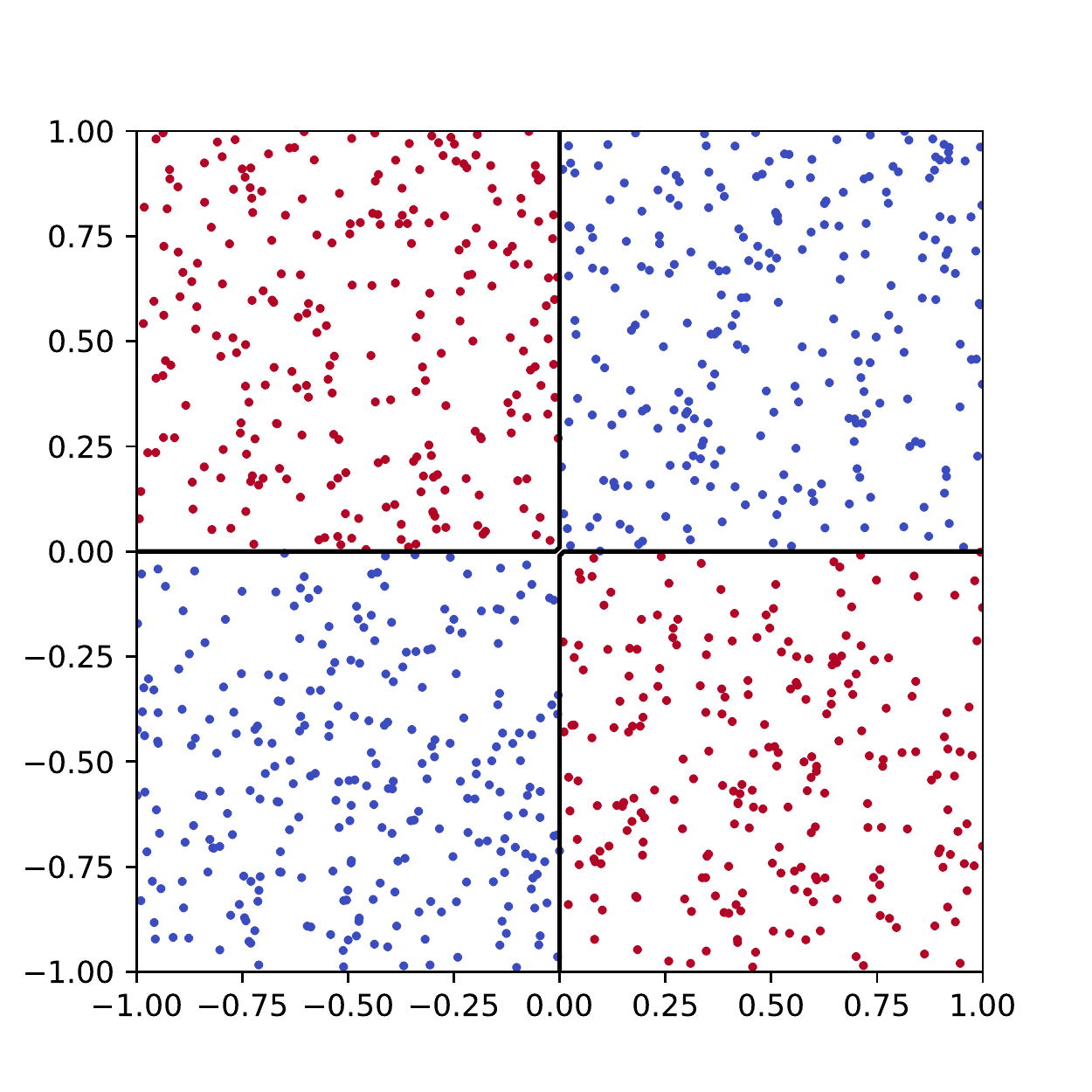}
 \includegraphics[width=0.49\textwidth]{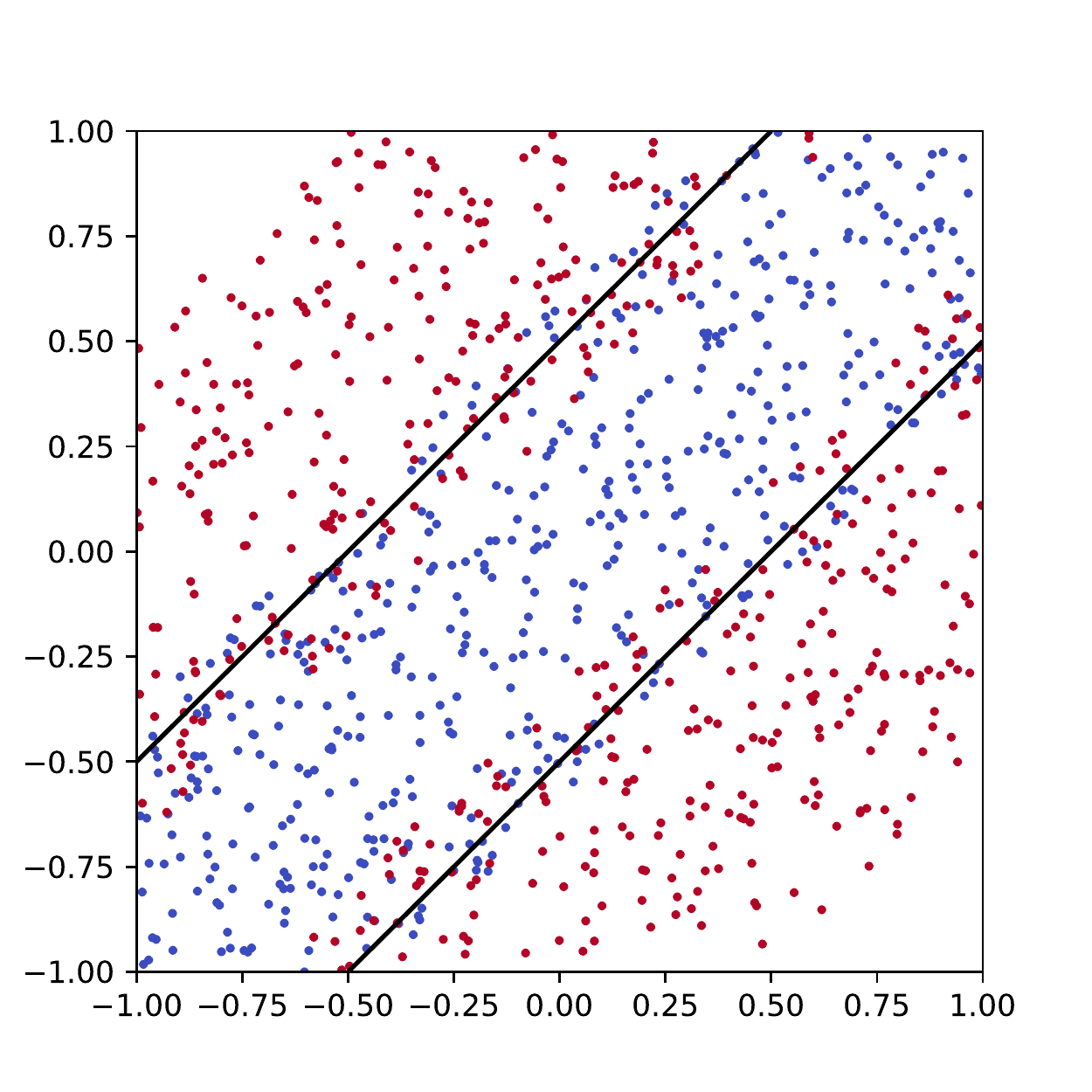}
 \includegraphics[width=0.49\textwidth]{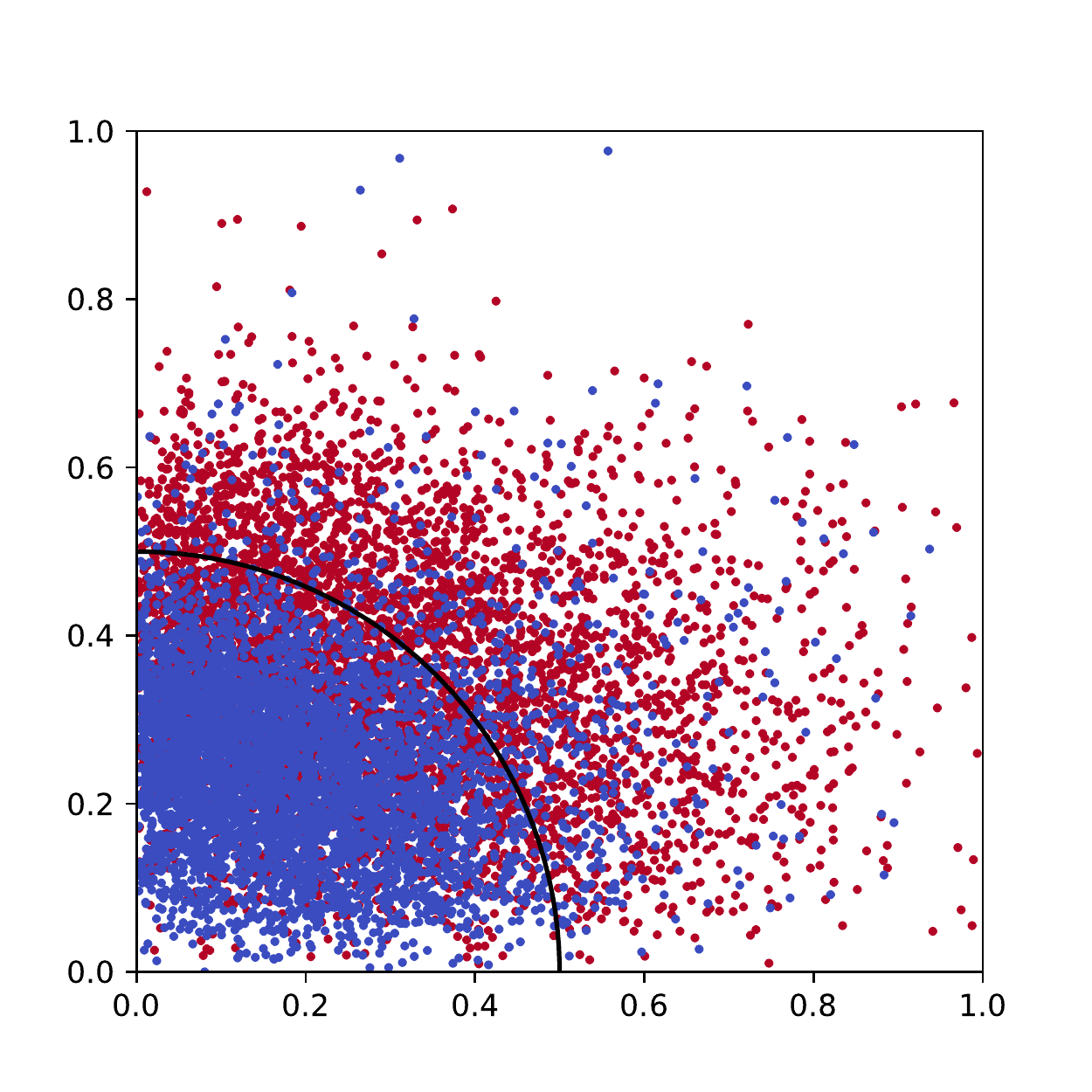}
 \caption{\label{fig:boundary}Decision boundary obtained with the quantum NN for each dataset.}
\end{figure*}

In practice, we find that the annealer performs best when the number of spins is well below its maximum capacity. Therefore in this study we pick a network with $N_f = N_h = 2$ and $N_b = 1$ for testing the performance of the training. The expressiveness of this network is limited, of course, but it is already enough to accurately classify samples in the following datasets (shown in Fig.~\ref{fig:boundary}):
\begin{itemize}
 \item \textbf{Circles.} Points in a noisy circle around the origin labeled $y = 1$, together with points in a blob inside the circle labeled $y = -1$.
 \item \textbf{Quadrants.} Uniformly-distributed points in a around the origin, labeled $y = -1$ if in the first or third quadrand, and $y = 1$ otherwise.
 \item \textbf{Bands.} Three partially-overlapping bands, parallel to the $x_2 = x_1$ direction, with labels $y = 1$, $y = -1$ and $y = 1$, respectively.
 \item \textbf{ttbar.} A set of events generated by simulations of proton-proton collisions at the LHC with final state containing 2 top quarks (see Ref.~\cite{Blance:2020nhl}). The label $y=1$ (the signal) indicates that the two tops are the decay products of a hypothetical new particle $Z'$~\cite{Altarelli:1989ff}. The label $y=-1$ (the background) means they arise from the known Standard Model physics. The features $x_0$ and $x_1$ correspond to the highest transverse momentum of a $b$-jet and the missing energy, respectively. Separating signal from background is relevant for experimental searches in this context~\cite{CMS:2012bti, ATLAS:2018rvc, ATLAS:2019npw}.
\end{itemize}
These datasets are best fitted with a smaller last-layer bias than $\pm 1$. We thus modify Eq.~\eqref{eq:weight-encoding} for $v_0$ so that it takes the values $-1/2$ or $0$ instead.
We use the D-Wave \texttt{Advantage\_system4.1} annealer to train the network with an annealing schedule given by
\begin{equation}
 s(t) = \left\{
  \begin{array}{ll}
   s_q \frac{t}{20} & \text{if } 0 \leq t < 20, \\
   s_q & \text{if } 20 \leq t < 80, \\
   s_q + (1 - s_q) \frac{t - 80}{20} & \text{if } 80 \leq t < 100,
  \end{array}\right.
\end{equation}
where $t$ is given in microseconds and $s_q = 0.2$. Since the more expensive part of the computation is the preparation of the annealer for the desired model, it is customary to run it several times once this is done to reduce noise in the output. We perform 300 runs and pick the final state with the least energy. In Fig.~\ref{fig:boundary} we show the decision boundary $Y(x) = 0$ obtained for each dataset using this procedure.

As a measure of the performance of the network, we use the area under the ROC curve. Its value is perfect (100\% area under ROC) for the \texttt{circles} and \texttt{quadrant} datasets, 92\% for the \texttt{bands}, and 78\% (close to the best attainable by other methods~\cite{Blance:2020nhl}) for \texttt{ttbar}.

\begin{figure}
 \centering
 \includegraphics[width=\linewidth]{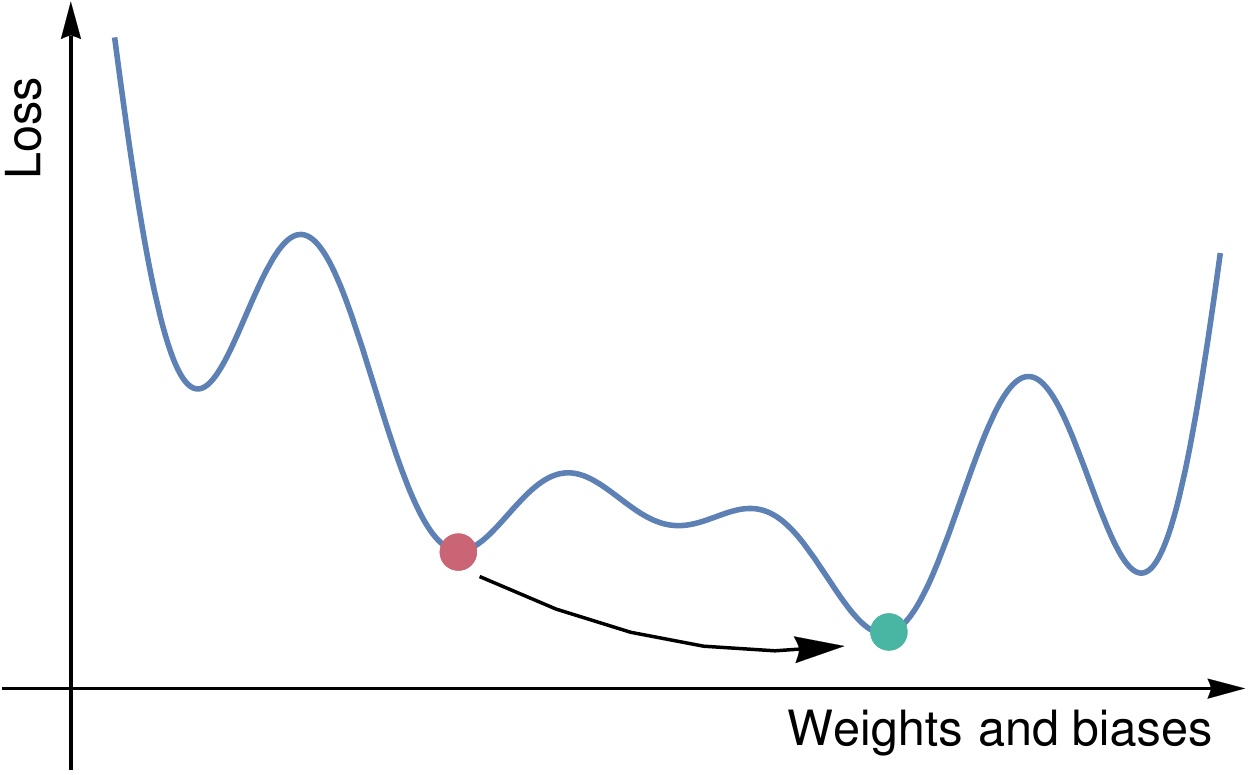}
 \caption{\label{fig:tunneling}Schematic representation of the loss function. Depending on the starting point and training method, classical training may end up in any local minima, such as the red one. Quantum training can tunnel from any of these minima to the global one, shown in green.}
\end{figure}

The advantage of quantum training is twofold. First, it can be performed in a single step instead of the incremental gradient-descent procedure used customarily in classical training.
Future annealers will be able to represent more extensive networks, which usually require large training times.
The quantum annealing procedure will be able to reduce them significantly.
Second, the quantum evolution can tunnel through barriers in the loss function to escape local minima, as depicted schematically in Fig.~\ref{fig:tunneling}.
This has been shown in Ref.~\cite{Abel:2021fpn}.
By contrast, classical algorithms must somehow surmount the barriers to find the global minimum.
Depending on the size of the barriers and on the hyperparameters of the training, classical methods may quickly become trapped in a local minimum and thus be unable to find the best solution to the problem.

To compare the effectiveness in finding the global minimum with classical algorithms, we trained a network with the same structure and activation function, using the \texttt{Adam} version of the (classical) gradient descent algorithm. When the network parameters are continuous, we obtained results with similar scores for the area under the ROC curve as with the quantum training. However, such a network has much greater freedom than the one implemented in the annealer, in which the parameters are only allowed to take two different values. To make the comparison more equitable, we mimicked this by adding
\begin{equation}
 \omega \sum_p (p - 1)^2 (p + 1)^2,\label{eq:2}
\end{equation}
to the classical loss function, where the sum is over all parameters (weights and biases) $p$. When $\omega$ is sufficiently large, the weights and biases are forced to adopt the same $p = \pm 1$ values available to the quantum annealer.

\begin{figure}
 \centering
 \includegraphics[width=\linewidth]{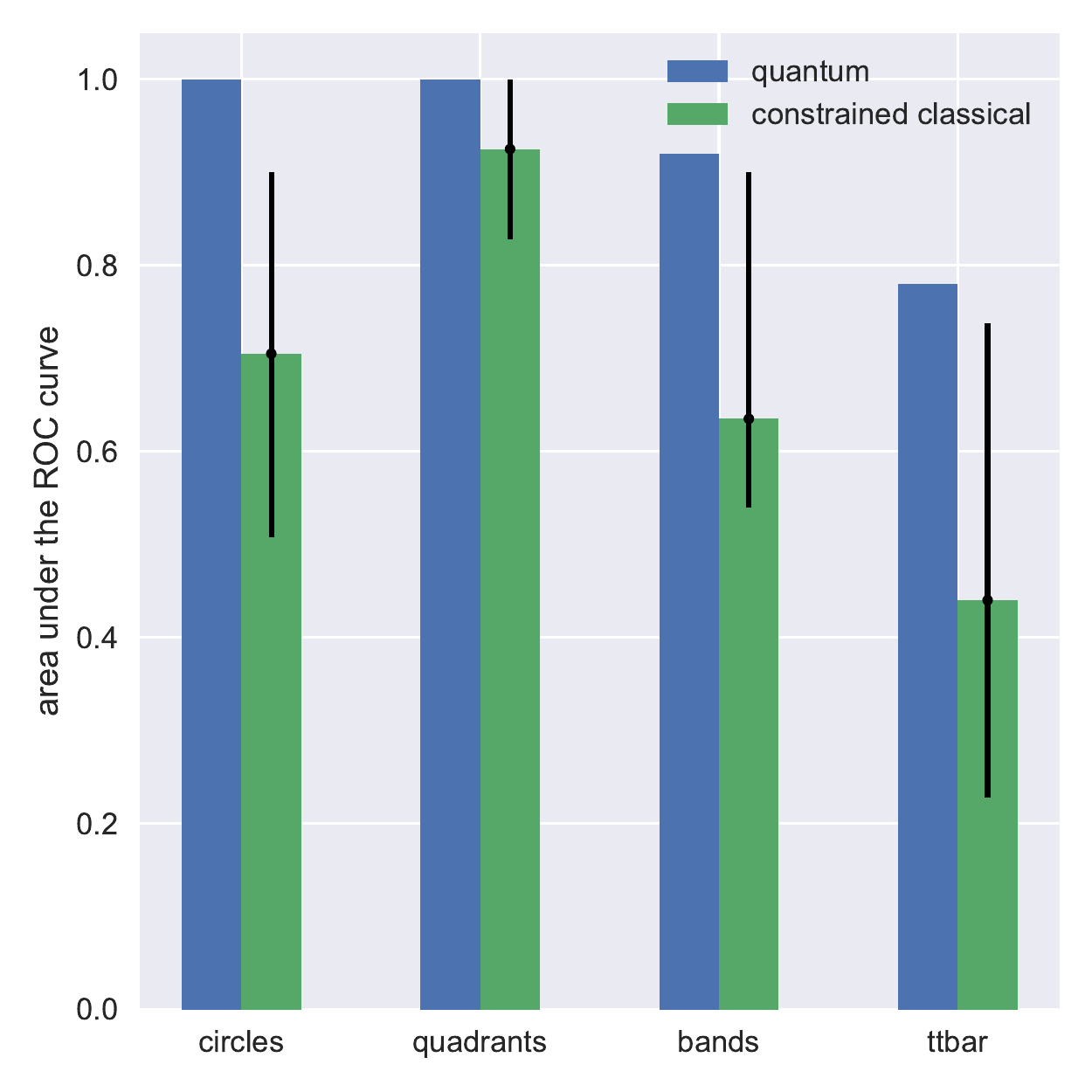}
 \caption{\label{fig:classical_quantum}Comparison of the results of the quantum NN to its classical analog. Each bar gives the median of 10 training runs. The error bars show the percentiles 20 and 80.}
\end{figure}

We find that classical training constrained in this way fails to reach the optimal solution most of the time with any of the four datasets, a practical manifestation of the situation sketched in Fig.~\ref{fig:tunneling}. A comparison between the results of the quantum and constrained classical training is shown in Fig.~\ref{fig:classical_quantum}. Apart from having a lower score on average, the classical training exhibits a significant variance in the results due to its becoming trapped in different local minima.

Finer discretization of the input data and network weights (that is, an encoding of each variable that uses more spins) will be made possible by future quantum annealers, as they are expected to have more qubits with more couplings between them. Moreover, the above comparison suggests that quantum annealers can operate and train neural networks, even with coarsely discretized weights and biases, where classical training would fail. This indicates that quantum training may quite rapidly be able to achieve better accuracy and do so more consistently than classical algorithms.\\

\noindent {\bf \underline{Acknowledgements}:}  We would like to thank Luca Nutricati for helpful discussions. S.A., J.C. and M.S. are supported by the STFC under grant ST/P001246/1.

\appendix

\section{Encoding binary higher-degree polynomials as quadratic ones}
\label{app:higher}

We prove here that, for any polynomial in binary spin variables $\sigma_\ell = \pm 1$, there exists a quadratic polynomial in an extended set of variables that includes auxiliary spins, such that the values of $\sigma_\ell$ at the global minima in the new system coincide with those in the old system. 

The proof is by induction. Consider a polynomial of degree $n$, in a system with $m$ spin variables $\sigma_{\ell=1...m}$. This can have many terms, but it is sufficient to focus on the terms of degree $n$. In general, we can write these terms in the polynomial as follows:
\begin{equation}
P_n (\lambda_\kappa)\equiv \sum_{W^n_\kappa} \lambda_\kappa W^n_\kappa  
\end{equation}
where $W^n_\kappa \equiv \sigma_{\ell_1}\sigma_{\ell_2}\ldots \sigma_{\ell_n} $ are all the words (that is monomials in which each spin can appear only once) of length $n$ that can be made from $m$ spins. Since $m\geq n$ there are ${}_mC_n$ of them labelled by $\kappa$ with couplings $\lambda_\kappa$ (some of which may be zero).

Consider a single pair of spins, say $\sigma_1$ and $\sigma_2$. Words containing the product $\sigma_1\sigma_2$ can be reduced by first converting to binary variables with $\sigma_\ell = 2\tau_\ell-1$, where $\tau_\ell=0,1$. Thus all such pairs translate as 
$$\sigma_1\sigma_2 \equiv 4\tau_1 \tau_1 -2\tau_1-2\tau_2+1~.$$ 
The linear and constant terms are contributions to the degree $n-1$ polynomial already, and the reduction task is therefore equivalent to reducing the $\tau_1\tau_2$ pair of binary variables. 

Consider adding to the polynomial a quadratic term involving the binary variables together with a new auxiliary variable $\tau_{12}$, which is of the form 
\begin{equation}
Q(\mbox{\small $\tau_{12};\tau_1,\tau_2$}) \, = \, \Lambda (\tau_1 \tau_2 - 2 \tau_{12} (\tau_1 + \tau_2) + 3 \tau_{12})~,
 \label{eq:constraint-Hamiltonian}
\end{equation}
where the overall coupling $\Lambda$ is chosen to be sufficiently large and positive. This Hamiltonian can of course be translated back into $\sigma_\ell$, however it is easier in QUBO format to check that it forces $\tau_{12} =\tau_1\tau_2$. Indeed if $\tau_1=\tau_2=0$ then 
$Q=3\Lambda \tau_{12}$ and if only one of $\tau_1$ or $\tau_2$ is zero then $Q=\Lambda \tau_{12}$. In both cases the minimum is at $\tau_{12} =0$ and $Q=0$. Meanwhile if $\tau_1=\tau_2=1$ then $Q=\Lambda (1-\tau_{12})$ and the minimum is at $\tau_{12} =1$ and $Q=0$.  

Thus adding $Q(\mbox{\small $\tau_{12};\tau_1,\tau_2$}) $ allows us to reduce all the words in $P_n$ that contain the product $\tau_1\tau_2$ by replacing the pair of binary variables with the single variable $\tau_{12}$. We note that 
the combined polynomial 
\begin{equation}
\widehat{P} (\tau_{12},\sigma_1,\sigma_2 \ldots) ~=~ {P}_{\widehat{12}} + Q(\mbox{\small $\tau_{12};\tau_1,\tau_2$}) 
\end{equation} 
in which $P_{\widehat{12}}$ is the original polynomial with the replacement $\tau_1\tau_2\rightarrow \tau_{12}$ in the $W^n_\kappa$ words, preserves the global minimum in $\sigma_1, \ldots \sigma_m$.

We then choose further pairs until the degrees of all the $W^n_\kappa$ words have been reduced, resulting in a degree $n-1$ polynomial with the same global minimum as the original. Note that in order to reduce all the words, the same spin may need to appear in more than one pair. This does not disrupt the proof, however, because the minima generated by $Q$ are degenerate and do not favour any value of $\tau_1$ or $\tau_2$ but simply fix $\tau_{12}$ to their product, which can then be replaced throughout. Having established that the degree of the polynomial can be reduced by one unit while preserving the global minimum, as required, the final step in the induction is to reduce the words of length $n=3$ to $n=2$, which follows in the same manner as for general $n$. \\

Finally let us verify that the reduction works correctly with a simple specific example, namely that provided by the Hamiltonian 
\begin{align}
H~&=~ \sigma_1\sigma_2\sigma_3~\nonumber \\
  &\equiv ~ 8 \tau_1\tau_2\tau _3 - 4 \tau_1\tau_2- 4 \tau_1\tau_3 - 4 \tau_2\tau_3 \nonumber \\ 
  & \qquad \qquad \qquad \qquad + 2 \tau_1 +2 \tau_2+ 2 \tau_3~,
\end{align}
where we drop the constant $-1$ in translating to the binary variables. This example is interesting because the degeneracy of global minima is lower than in the generic case: there are only 4 solutions to $\sigma_1\sigma_2\sigma_3=-1$ (which corresponds in binary language to any one of the $\tau_\ell$ being zero, or all of them), as opposed to the seven solutions to $\tau_1\tau_2\tau _3 =0$. One might therefore be concerned that new degenerate minima may appear, but this is not the case. As usual we now reduce the trilinear term by trading $\tau_1 \tau_2$ for an auxilliary binary $\tau_{12}$ by adding the Hamiltonian $Q(\mbox{\small $\tau_{12};\tau_1,\tau_2$}) $, and replacing the pair in $H$: that is the new QUBO Hamiltonian becomes 
\begin{align}
H~& = ~ Q(\mbox{\small $\tau_{12};\tau_1,\tau_2$})~+~ 
8 \tau_{12} \tau _3 \nonumber \\
& ~~\qquad - 4 \tau_{12} - 4 \tau_1\tau_3 - 4 \tau_2\tau_3
 \nonumber \\ 
 & ~~~~\qquad\qquad + 2 \tau_1 +2 \tau_2+ 2 \tau_3~. 
\end{align}
It is easy to check that provided $\Lambda>2$ the original 4 degenerate solutions hold in the new combined Hamiltonian as required.

\bibliographystyle{inspire}
\bibliography{references,referencesSAMS}

\end{document}